\documentstyle[11pt,newpasp,epsf]{article}
\def\edcomment#1{\iffalse\marginpar{\raggedright\sl#1\/}\else\relax\fi}
\reversemarginpar

\begin{document}
\title{The Instability Strip of M3}
\author{G. \'A. Bakos, J. Jurcsik}
\affil{Konkoly Observatory, P.O. Box 67,
H-1525 Budapest XII, Hungary}

\begin{abstract}
We present new multicolour CCD photometry of the central part of the
globular cluster M3, mapping the precise position of $\sim 120$ RR Lyrae
stars (RRab, RRd, RRc) on the horizontal branch (HB). The location of the
double-mode variables (RRd) is in perfect agreement with recent theoretical
results. We find a significant internal spread of metallicity amongst the
RRab variables.
\end{abstract}

\noindent CCD observations were carried out using the 1-m RCC telescope at
the Piszk\'es\-tet\H{o} Mountain Station of Konkoly Observatory. Standard
crowded field photometry with {\sc allframe} (Stetson, 1994) yielded
light curves with complete phase coverage for $\sim$ 120 RR Lyrae stars, as
well as precise magnitudes for the non-variable stars of the HB. A detailed
discussion of observations and data reduction will be given in a forthcoming
paper (Bakos et al.~1999). 

\begin{figure}[!h]
\plotfiddle{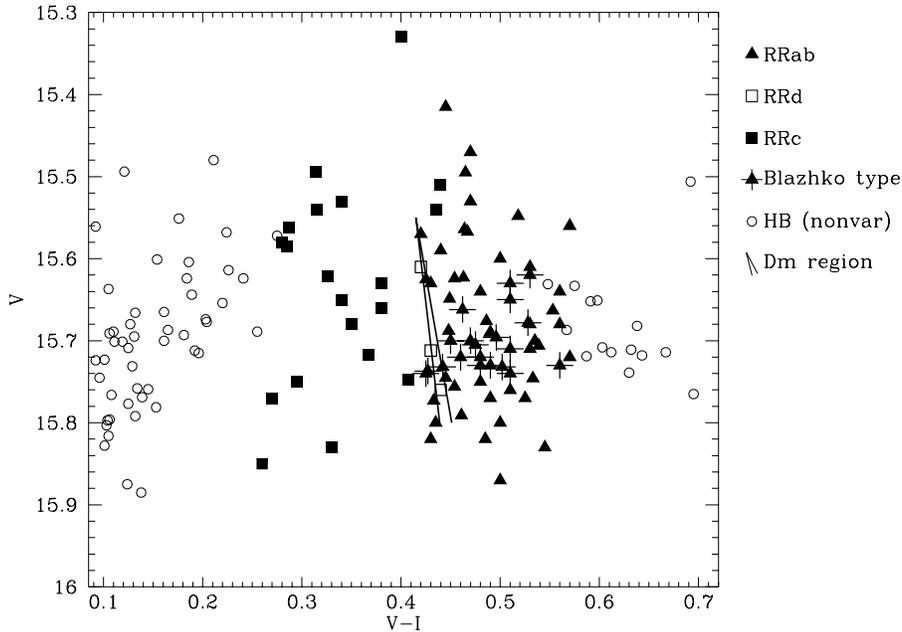}{8.5cm}{-90}{45}{45}{-180}{260}
\caption{The horizontal branch of M3 with the observed RR Lyrae stars
plotted.  The double-mode regime (Szab\'o et al. 2000) is also shown.}
\end{figure}

The colour--magnitude diagram of the HB was constructed using 
the \linebreak weighted average magnitude of the best $\sim 15$ frames 
for the non-variable stars, while the magnitudes and colours for all known
variables were derived from optimal-order Fourier fits to the light curves.
Fig. 1 reveals that the HB has a considerable width, notably exceeding
the photometric errors (which are estimated to be less than 0.03 mag). This
can either be interpreted as an unusual stellar evolution scenario of the
HB, or a significant internal spread in the metallicity. 

Blazhko variables
have markedly smaller scatter in $V$; however, this might be due to small
number statistics. The RRab and RRc populations are well-separated, with
minimal overlap. The RRd regime is relatively narrow in colour, though it
should be noted that only three of the five known RRd variables are
within our observed fields.  The position of the RRd stars is in perfect agreement 
with the theoretical prediction of Szab\'o et al. (2000) using double-mode 
RR~Lyrae models with $Z=0.001$, and $M_{RR}=0.65\,M_{\odot}$. Bolometric 
corrections
and temperature--colour transformations were derived by employing adequate
grids of stellar atmosphere models from Castelli, Gratton, \& Kurucz 
(1997). The
observations precisely mapped the morphology of various parts of the HB,
which can serve as a check on further theoretical calculations.

Metallicity and reddening of the single-mode RRab stars were determined by
applying relations between their physical and light curve parameters
(Jurcsik \& Kov\'acs 1996, Kov\'acs \& Jurcsik 1997). The average
metallicity of the cluster turns out to be $\rm [Fe/H]=-1.37\pm0.17$
(std.~dev.), where the large scatter can neither be ascribed to photometric
errors, nor to the inaccuracy in the aforementioned methods. The almost
negligible reddening value of $E(V-I)=0.00\pm0.02$ obtained from comparison
of the observed and calculated colours agrees well with previous estimates.

\begin{acknowledgements}
This work was supported in part by OTKA T-024022 and T-030954. 
\end{acknowledgements}

\end{document}